\documentclass[useAMS,usenatbib,onecolumn]{mn2e}
\newcommand{\Div}{\vec{\nabla} \cdot }
\newcommand{\Rot}{\vec{\nabla} \times }
\newcommand{\Grad}[1]{\vec{\nabla} {#1} }
\newcommand{\EV}{\vec{E}}
\newcommand{\BV}{\vec{B}}
\newcommand{\LD}{\mathcal{L}}
\usepackage{graphicx}
\title[Stationary electromagnetic fields]
 {Stationary electromagnetic fields in the
exterior of a slowly rotating relativistic star: a description beyond the
low-frequency approximation}
\author[Y. Kojima, N. Matsunaga and T. Okita ]
{Yasufumi Kojima\thanks{E-mail:
kojima@theo.phys.sci.hiroshima-u.ac.jp},
Norihito Matsunaga 
and Taishi Okita\\
Department of Physics, 
Hiroshima University,
Higashi-Hiroshima 739-8526, Japan}

\begin{document}


\pagerange{\pageref{firstpage}--\pageref{lastpage}} \pubyear{2003}

\maketitle

\label{firstpage}

\begin{abstract}
We investigate the electromagnetic fields in the vacuum exterior of a
rotating relativistic star endowed with a magnetic dipole moment,
and with the stellar surface behaving as a perfect conductor.
While the stellar rotation is treated in the slow-approximation 
of general relativity, we do not restrict our
attention to slowly rotating electromagnetic fields, 
and take our analysis beyond the low-frequency approximation 
considered so far.   When the dipole moment is misaligned 
with the rotational axis, our approach does
not yield analytic solutions as in Rezzolla et al. (2001a, b), but
determines the properties of the electromagnetic fields
approximately and semi-analytically,
by computing the coefficients of simple expressions
for the fields through the numerical solution of two partial differential
equations. Because our approach provides a solution which is in principle
valid throughout space, we can evaluate the accuracy and/or invalidity
of previously known analytic expressions at different distances
from the stellar surface.
Overall, the solutions found in this way represent an efficient way
of bridging in a single semi-analytic formalism the strongly 
relativistic (Rezzolla et al. 2001a, b) and the asymptotic
regimes (Deutsch 1955) for which analytic solutions have been found.
\end{abstract}


\begin{keywords}
relativity -- stars: magnetic fields -- stars: 
neutron -- stars: rotation
\end{keywords}

\section{Introduction }

	It is well known that magnetic fields often play a key role in
astrophysics and  an ample discussion is given, for instance, in \cite{bp1,
bz1}. In the limit of very high electrical conductivity, 
the magnetic flux is ``frozen'' into the fluid and increases as a result of
fluid compressions. In this case the the typical value of the magnetic
field $B$ can be related to the rest-mass density $ \rho$ through the
simple relation $ B \propto \rho ^{2/3}$, so that very strong magnetic
field may be produced in compact stars, where the gravity is also strong.
Some observations indicate that the magnetic field in young neutron stars is
of order $10^{12}$ G.  Recently, exceptionally stronger cases with
$10^{14}-10^{15}$ G are suggested in the phenomena with soft gamma
repeaters and anomalous X-ray pulsars \citep{ms1, ok1, ok2}.
Very intense magnetic fields may be produced through a dynamo action
\citep{td1} and the neutron stars with such magnetic fields are usually
referred to as magnetars. The ideas behind the existence and formation of
magnetars are gaining support from the observations and a
number of authors have discussed the mechanisms generating these magnetic
fields and their astrophysical relevance (see, for instance, \citet{td2,
td3}).
The combined effects of strong gravity and intense electromagnetic 
fields may be
crucial in some astrophysical objects and it is therefore important to
consider the electromagnetic field in curved spacetime produced by strong
gravity. The study is important in clarifying the general relativistic
effects on the electromagnetic fields, and also indispensable in
constructing realistic models.
%

   A number of works have been so far published concerning the effects of
strong gravity on the electromagnetic fields. The magnetic field in
principle affects the spacetime curvature through the Einstein equations,
but the actual deformation at the surface is of order $10^{-5} (B/10^{15}
{\rm G} )^2 $ even for magnetars \citep{k1, k2}.
In most astrophysical cases, therefore, the deformation may be neglected,
and general relativistic effects can be evaluated by solving the Maxwell
equations in fixed but curved spacetime.  Several attempts are done along
this line of thought.  Stationary electromagnetic fields have been
considered both in Schwarzschild spacetimes (e.g. \citet{g1, a1, p1}) and
in slowly rotating spacetimes (e.g. \citet{m1, m2, k3}).

	More recently, \citet{r1, r2} and \citet{z1} have extensively
studied the interior and and vacuum exterior electromagnetic fields
produced by a rotating dipole moment comoving with the star. Their
analysis is fully general relativistic and has provided compact and
analytic expressions for both the cases in which the magnetic dipole is
aligned or not with the stellar rotational axis.
In these works, the star of mass $M$ and radius $R$ is assumed to be
rotating at an angular velocity $\Omega_*$ much smaller than the maximum
angular velocity, i.e. the Keplerian angular $\Omega_{_{K}} \simeq
\sqrt{GM/R^3}$, so that a {\it slow-rotation approximation} can be used
satisfactorily to describe the general relativistic corrections
introduced by the rotation of the spacetime. Furthermore, the
electromagnetic fields are assumed to be comoving with the star at all
times, thus possessing an intrinsic angular velocity $\Omega=\Omega_* \ll
\Omega_{_{K}}$; as a result, the expressions of Rezzolla et al. (2001a,
b) are expected to be accurate in the vicinity of the star, where the
general relativistic corrections are stronger and the intrinsic
velocities smaller. We will refer to this as to the {\it low-frequency
approximation} in the Maxwell equations.

	While the slow-rotation approximation is a good one except for
extremely rapidly rotating stars, the low-frequency approximation may
have some limitations. The first one is that the expressions of the
electromagnetic fields derived in this case cannot be accurate at larger
radii, (i.e. in the vicinity and beyond the light cylinder), where they
should manifest a wave-like nature. The second limitation may arise in
the case in which the electromagnetic fields are not comoving with the
star and, more specifically, in the case in which $\Omega \gg \Omega_*$.

	In this paper we circumvent these limitations by considering the
general relativistic electromagnetic fields exterior to a rotating
neutron star in the slow-rotation approximation but {\it not} in the
low-frequency approximation. We will not discuss here the physical
conditions that may lead to electromagnetic fields with $\Omega \gg
\Omega_*$, nor whether this is likely to happen for realistic
astrophysical sources (indeed this condition may be difficult to occur in
practice); rather we will make this our working hypothesis. As a result,
we here consider the electromagnetic fields in the vacuum exterior of a
relativistic star slowly rotating at frequency $\Omega_* \ll
\Omega_{_{K}}$ and endowed with a magnetic dipole moment rotating at a
frequency 
(either $\Omega = \Omega_*$ or $\Omega \gg \Omega_*$ ).
Furthermore, the stellar surface is
assumed to be a perfect conductor, thus inducing a quadrupolar electric
field \citep{d1,r1}.  Our approach differs from previous ones both in the
underlying assumptions (i.e. high frequency electromagnetic fields) but
also in the method. In particular, we do not provide analytic expressions
(as in Rezzolla et al. 2001a, b) but determine the properties of the
electromagnetic fields through a semi-analytical approach in which simple
expressions for the electromagnetic fields need to be completed through
the calculation of coefficients coming from the numerical solution of two
partial differential equations. Because the approach is based on a series
expansion, the solution and its accuracy depends on the number of terms
considered. On the other hand, our approach yields solutions that are
valid in the whole spatial domain and this allows us to compare our
results with the analytic expressions of Rezzolla et al. (2001a, b) and
of Deutsch (1955), estimating the accuracy of these analytic solutions
in those regimes where they are not expected to be valid.

	The paper is organized as follows.
In Section 2 we summarize the Maxwell equations in vacuum with stationary
condition. The equations are reduced to a set of two partial differential
equations among two scalar potentials. Here, we discuss how it is
possible to to look for simple Deutsch-type solution which are
approximate solution of the general relativistic Maxwell equations in a
slowly rotating spacetime.
In Section 3, we show the explicit calculations in the case in which the
magnetic dipole is parallel to the rotational axis, while
in Section 4, we consider the case when the dipole is perpendicular.
Because of the linearity of the Maxwell equations, the general case of an
oblique rotator can be constructed from the superposition of these two
particular cases.
Concluding remarks are finally presented in Section 4. Hereafter we will
use units in which $c=G=1$.

\section{ Stationary Electromagnetic field}

%
We will consider the solution to the Maxwell equations 
in the fixed background spacetime of a rotating relativistic star whose
generic axially symmetric line element is given by
\begin{eqnarray}
\label{ds2}
 ds^2 &=&
  - \alpha ^2 dt^2  
+ \gamma _{ij}(dx^i + \beta ^i _{*} dt)(dx^j + \beta ^j _{*} dt)
\\  
 &=&
 - \alpha ^2 dt^2  
 +e^{2\psi} (d\phi - \omega  dt )^2
 +e^{2\mu} dr^2  +e^{2 \eta} d\theta ^2 .
\end{eqnarray}
In practice and except for rapidly rotating stars, it may be sufficient
to take into account the first-order rotational effect of the metric only
(see, for instance \citep{z1}) so that the metric functions exterior to
the slowly rotating star are reduced to the form
\begin{eqnarray}
\label{ds2_2}
&&
 \alpha \to \left( 1- \frac{2M}{r} \right)^{1/2} , ~~
 e^{ \mu } \to \left( 1- \frac{2M}{r} \right)^{-1/2}, ~~
 e^{\eta } \to r, ~~
%
 e^{ \psi } \to r \sin \theta, ~~
 \omega   \to    \frac{2J}{r^3} ,
\end{eqnarray}
where $M$ and $J$ are the mass and angular momentum of the central star.
In this paper, the electromagnetic fields are denoted by the values
measured by ``zero angular momentum observers'', i.e. ZAMOs \citep{b1},
and their components projected to a locally orthonormal tetrad frame are
expressed as $( E_r, E_\theta, E_\phi)$ and $( B_r, B_\theta, B_\phi)$ .
These are sometimes expressed with hatted indices to distinguish them
from coordinate values (see, for instance, Rezzolla et al. 2001a, b);
however, because we will always refer only to the electromagnetic fields
as measured by  ZAMOs, we will omit here the hatted indices.

	The electric and magnetic fields in vacuum satisfy the Maxwell
equations in this curved spacetime (e.g. \cite{t1}).
\begin{eqnarray}
\label{eqn.dvb}
&&   \Div \BV  = 0,
\\
\label{eqn.dve}
&&     \Div \EV  = 0,
%
\\
\label{eqn.rte}  
&& 
   \partial_t  \BV - \LD_{\vec \beta _* }  \BV   = 
 - \Rot (\alpha \EV) ,
\\
\label{eqn.rtb}  
&& 
   \partial_t  \EV - \LD_{\vec \beta _* } \EV   = 
 \Rot (\alpha \BV)  ,
%
\end{eqnarray}
where $\LD_{\vec \beta _*}$ is the Lie derivative along the vector field
${\vec \beta _*}\equiv - \omega \partial _\phi $ and is defined for any
vector $ {\vec A} $ as
\begin{equation}
\LD_{\vec \beta _*} {\vec A}
=  ({\vec \beta _*} \cdot \Grad ){\vec A} - 
   ({\vec A} \cdot \Grad ) {\vec \beta _*} .
\end{equation}

%
Clearly, in the frame rotating with angular velocity $ {\vec \beta }
\equiv \Omega \partial _\phi$, the electromagnetic fields are independent
of time and these are the electromagnetic fields we will consider
hereafter.
Note that the angular velocity $ \Omega $ may differ from that of the
star $\Omega _*  $, which 
is related to the spin angular momentum $ J $ and the inertial moment
$I$  through the relation $\Omega _* = J/I $
in the slowly rotating spacetime. 
For example, the electromagnetic fields can be produced by a periodic
motion with a certain angular frequency $ \Omega $ on or outside the
star.
Furthermore, and as mentioned in the Introduction, 
we do not assume that the frequency $
\Omega $ is small, unlike in previous work.
The stationarity of a scalar function $f$ 
is equivalent to expressing that $f$ should depend on
the time $t$ and on the azimuthal angle $\phi$ only in a combination of
the type $ \phi ' = \phi - \Omega t $. In the locally flat
four-dimensional spacetime in which the ZAMOs make their measurements,
this condition of stationarity for the function $f$ can be expressed by
requiring that the function $f$ is solution of the equation: 
$\partial_tf + {\vec \beta } \cdot {\vec \nabla} f =0$, 
i.e. that the function is a
constant along the direction $(1, {\vec \beta} )$, in the
four-dimensional reference frame of the ZAMOs. In the case of the
electromagnetic vector fields $ {\vec E} $ and $ {\vec B}$, the
mathematical condition for stationarity needs to be expressed through the
Lie derivative as
%
%
\begin{equation}
 \partial_t  \EV  + \LD_{\vec \beta } \EV = 0,
~~~~~~
 \partial_t  \BV  + \LD_{\vec \beta  } \BV = 0.
\end{equation}

%
Using these conditions in eqs.(\ref{eqn.rte}) and (\ref{eqn.rtb}), some
combinations of $\EV$ and $\BV$ are expressed by gradient of two scalar
potentials, $H$ and $G$.  After solving for $\EV$ and $\BV$, we have
\begin{equation}
\EV  = - \frac{1}{\alpha \Lambda }
 \left[ 
    e^{- \mu }  G , _{r}
 +  \varpi e^{\psi -\eta}  \alpha ^{-1}  H , _{\theta} , ~
   e^{ -\eta } G , _{\theta}
- \varpi e^{ \psi -\mu} \alpha ^{-1}  H , _{r} , ~
 e^{-\psi} \Lambda  G , _{\phi}
\right] ,
\label{evct}
\end{equation} 
\begin{equation}
\BV  = - \frac{1}{\alpha \Lambda }
 \left[
   e^{- \mu }  H,  _{r}
 -  \varpi e^{\psi -\eta}  \alpha^{-1} G , _{\theta} , ~
 e^{ -\eta } H , _{\theta}
+  \varpi e^{\psi -\mu}  \alpha^{-1} G,  _{r}  , ~
  e^{-\psi} \Lambda  H , _{\phi}
\right] ,
\label{bvct}
\end{equation} 
where a comma denotes a partial derivative with respect to the coordinate
$j (=r, \theta, \phi)$ and
\begin{equation}
\Lambda = 1- 
 \varpi^2 e^ {2\psi} \alpha^{-2} ,
\end{equation} 
\begin{equation}
\varpi = \Omega - \omega. 
\end{equation} 
From eqs.(\ref{eqn.dvb}) and (\ref{eqn.dve}), we have a pair of 
second-order partial differential equations
\begin{eqnarray}
&&
\Lambda 
 \left[
\alpha  \left(   e^{- \mu +\eta + \psi } \alpha^{-1}
         G _{,r} \right) _{,r} 
+
\alpha  \left(   e^{\mu -\eta +\psi } \alpha^{-1}
         G _{,\theta} \right) _{,\theta} 
+ e^{\mu + \eta -\psi} 
\Lambda  
         G _{,\phi, \phi} 
\right]
 \nonumber
\\
&& ~~~~~~
 -e^{- \mu +\eta + \psi }  \Lambda _{,r}   G _{,r}
-e^{\mu -\eta +\psi }   \Lambda _{,\theta}   G _{,\theta}
+ K_r H _{,\theta} -K_\theta H _{,r} 
=0,
\label{twop1}
\\
&&
\Lambda 
 \left[
\alpha  \left(  e^{- \mu +\eta + \psi }  \alpha^{-1}
         H _{,r} \right) _{,r} 
+
\alpha  \left(  e^{\mu -\eta +\psi }  \alpha^{-1} 
         H _{,\theta} \right) _{,\theta} 
+ e^{\mu + \eta -\psi} 
\Lambda  
         H _{,\phi, \phi} 
\right]
 \nonumber
 \\
&& ~~~~~~
 - e^{- \mu +\eta + \psi }  \Lambda _{,r}   H _{,r}
 - e^{\mu -\eta +\psi }  \Lambda  _{,\theta}   H _{,\theta}
- K_r G _{,\theta} + K_\theta G _{,r} = 0 ,
\label{twop2}
\end{eqnarray}
%
where
\begin{equation}
K_j  =
\alpha  
  \left(  \frac{ \varpi e^{ 2 \psi } }{ \alpha ^2 } 
  \right) _{,j}   
+ \frac{ \varpi ^2  e^{ 4 \psi } }{ \alpha ^3 } 
   \left( \varpi \right) _{,j} 
~~~~~~~~~~~
({\rm for}~~j=r,\theta).
\end{equation}
These equations (\ref{twop1}) and (\ref{twop2}) are symmetric under the
transformation between ${\vec E}$ and ${\vec B}$, i.e. ${\vec E} \to
{\vec B}$, by changing $ G \to H $ and $ H \to -G $.
Note that this is 
a generic property of the source-free Maxwell equations.

\section{ Axially Symmetric  Case }
%
We now look for solutions of eqs.(\ref{twop1}) and (\ref{twop2}) in
the slowly rotating spacetime.
Note that these equations have a singularity at the light cylinder $
\Lambda = 0$, so that a special care should be paid when performing a
numerical integration.
We now expand the electromagnetic fields using a series expansion with
respect to the angular part which is particularly advantageous for the
imposition of regularity conditions.
Since the electric
and magnetic fields are respectively polar and axial vectors with respect
to parity transformation, we expand them through two scalar potentials
$G$ and $H$, which have parity of $+1$ and $ -1$, respectively\footnote{
The function $H$ is not a scalar one, but rather a "pseudo-scalar".}.
Therefore, for an axially symmetric system, these functions are
respectively expanded in terms of the spherical harmonics $Y_{2l} ^0 $
and $Y_{2l+1}^0 $. Note that the following forms are equivalent to the
spherical harmonics expansion, with the difference that the regularity
condition can be imposed more easily.
%
%
\begin{equation}
 G(r, \theta) = 
\sum _{n=0} ^{\infty}  g_n(r) (\sin \theta)^{2n} ,
\label{expndm0G}
\end{equation}
\begin{equation}
 H(r, \theta) =\cos \theta \left( \sum _{n=0} ^{\infty}
 h_n(r) (\sin \theta)^{2n}
\right) .
\label{expndm0H}
\end{equation}
The electromagnetic fields are described as
\begin{eqnarray}
B_r &=& 
\frac{ - \cos \theta } { \Lambda }
\left[   \sum _{n=0} ^{\infty}
\left(  h_n ' -\frac{2n  \varpi}{\alpha^2 }  g_{n} \right)
  (\sin \theta)^{2n}
\right] ,
\label{axbr}
\\
B_\theta &=& 
 \frac{ -\sin \theta} {  \alpha r \Lambda }
\left[ \sum _{n=0} ^{\infty} \left\{
\varpi r^2  g_{n} ' +  2(n+1) h_{n+1}  -(2n+1) h_{n} 
\right\} (\sin \theta)^{2n} \right]  ,
\label{axbt}
\\
E_r &=& 
  \frac{ -1 } { \Lambda }
    \left[  g_0 ' + 
     \sum _{n=1} ^{\infty} \left\{
 g_n ' + \frac{ \varpi }{\alpha^2 } 
(  2n h_{n}  -(2n-1) h_{n-1} ) \right\}  (\sin \theta)^{2n}
    \right] ,
\label{axer}
\\
E_\theta &=& 
\frac{  \sin \theta \cos \theta} {  \alpha r \Lambda }
 \left[ \sum _{n=0} ^{\infty} \left\{
  \varpi r^2 h_n ' -2(n+1) g_{n+1} 
 \right\} (\sin \theta)^{2n} \right]  ,
\label{axet}
\end{eqnarray}
where a prime $'$ denotes a radial derivative and
\begin{equation}
\Lambda =  1- b^2 \sin^2 \theta
= 1- \left( 
\frac{ \varpi r }{ \alpha } \right) ^2   \sin^2 \theta \ .
\end{equation}
Note also that the expansions (\ref{expndm0G}) and (\ref{expndm0H}) are not
necessarily valid in whole space.  The denominators of eqs.(\ref{axbr}) -
(\ref{axet}) become zero at the light cylinder, which is located at $
b(r) \sin \theta =1$.
Taking the limit of $ \sin \theta \to 1/b(r) $ in the numerators, we have
two regularity conditions among the radial functions, $ g_{n} $ and $
h_{n} $
\begin{equation}
\sum _{n=0} ^{\infty}
\left\{ 
 g_{n} ' + \frac{ \varpi }{\alpha^2 } 
\left(
  2n   - \frac{2n+1 }{ b^2} \right) h_{n} 
\right\} b^{-2n}  = 0,
\label{regm0p1}
\end{equation}
and
\begin{equation}
\sum _{n=0} ^{\infty}
\left(  h_n ' -\frac{2n  \varpi}{\alpha^2 }  g_{n} \right)
b^{-2n}
=0.
\label{regm0p2}
\end{equation}
These relations should be satisfied for
the region $ 1/b \le  1$,  
which is equivalent to $ r \ge r_c$, where
the radius $r_c$  is defined by
$b(r_c)=1$ and, of course, $r_c$  is reduced to $ 1/\Omega $
in the flat spacetime.
Note that it is not necessary for the radial functions to 
satisfy the conditions (\ref{regm0p1}),(\ref{regm0p2})
in the inner region $r < r_c $.
It may be possible to construct a global solution 
by smoothly matching the radial functions across  the
radius $ r_c $,
which are solved 
with  and without the constraints (\ref{regm0p1}),(\ref{regm0p2})
on either side of $r_c$. 
We however look for the solutions that satisfy  (at least approximately)
  the constraints (\ref{regm0p1}), (\ref{regm0p2})
in the whole space exterior to the stellar surface.
This method may restrict the class of solutions, but
avoids to locate and handle a matching at $r_c$. 
If no solution can be found satisfying the
constraints (\ref{regm0p1}), (\ref{regm0p2}) in in entire region outside
the star, we will then look for solutions under weaker constraints.

We consider the electromagnetic field
described or approximated 
by a finite number of functions
$g_n (r)$ and $h_n (r) $.
For example, 
the field by rotating magnetic dipole
in a flat spacetime can be described by 
$    g_n ,  h_n (n=0,1) .$
See Appendix for the results in the flat case
for comparison.
It is natural to approximate the  electromagnetic fields 
in a similar fashion also in the case of a slowly rotating spacetime.

The regularity conditions for the non-vanishing components $ g_n , h_n
(n=0,1) $ are given by
\begin{equation}
 0 ={\cal X}
\equiv 
g_1 '  +b^2 g_0 ' - \frac{\varpi}{ \alpha ^2} h_0 
- \frac{1}{\varpi  r^2 } 
\left(3 - 2 b^2  \right) h_1 
\label{eq0g1}
\end{equation}
and
\begin{equation}
 0 = {\cal Y} 
\equiv
h_1 '  +b^2 h_0 ' - \frac{2\varpi}{ \alpha ^2} g_1 .
\label{eq0h1}
\end{equation}
Using these conditions in eqs.(\ref{axbr})-(\ref{axet}),
we have
\begin{equation}
B_r = -h_0 ' \cos \theta ,
\end{equation}
\begin{equation}
B_\theta
 = \frac{1}{\alpha r}
\left(  h_0  - 2 h_1 - \varpi r^2 g_0 ' \right)
\sin \theta  ,
%
\end{equation}
\begin{equation}
E_r = 
- g_0 ' 
-\frac{3}{\varpi  r^2} h_1 \sin ^2 \theta 
=
- \left( g_0 ' + \frac{2}{\varpi  r^2} h_1  \right)
+ \frac{2}{\varpi  r^2} h_1 P_2 (\cos \theta),
\end{equation}
\begin{equation}
E_\theta  = \frac{1}{\alpha r}
 (\varpi  r^2 h_0 '-2 g_1) \sin \theta \cos \theta  .
%
\end{equation}
The electric field is given by a mixture of monopolar and quadrupolar
terms. In particular, when the total electric charge is zero, the
monopolar part vanishes, with the quadrupolar one being the one induced
by rotation. This condition can be written as
\begin{equation}
 g_0 ' + \frac{2}{\varpi  r^2} h_1 =0.
\label{nomople1}
\end{equation}

The divergence of the magnetic field yields
\begin{equation}
\Div \BV  = - \alpha {\cal H } \cos \theta  , 
\end{equation}
where 
\begin{equation}
{\cal H }
= h_0 ''  + \frac{2}{r} h_0 '
+ \frac{ 2\varpi  }{\alpha^2} g_0 ' 
+\frac{2}{\alpha^2 r^2} ( 2 h_1 - h_0) . 
\label{eq0H00}
\end{equation}
Using the condition (\ref{nomople1}) in eq.(\ref{eq0H00}), we have a
second order differential equation of $ h_0$ only.  Enforcing that the
magnetic field is divergence-free is equivalent to setting $ {\cal H} =
0$ so that equation (\ref{eq0H00}) has solution
\begin{equation}
 h_0 = \frac{3 \mu}{4M^3}
\left[r \alpha^2 \ln \alpha ^2  - \frac{2M^2}{r} +2M
\right],
\label{solh0}
\end{equation}
where $ \mu $ denotes the magnetic dipole moment.  This relativistic
solution for the magnetic field is well known in the literature
\citep{g1, a1, p1}.

We now express the divergence of the electric field as 
\begin{equation}
\Div \EV  = - \alpha {\cal G }  + {\cal Q } \sin ^2 \theta ,
\end{equation}
where 
\begin{equation}
{\cal G }
=  g_0 ''  + \frac{2}{r} g_0 '
- \frac{ 2\varpi  }{\alpha^2} h_0 ' 
+\frac{4}{\alpha^2 r^2} g_1 ,
\label{eq0G00}
\end{equation}
and
\begin{equation}
{\cal Q } =
\frac{ 3 \alpha  }{ \varpi r^2} 
\left(
\frac{ \varpi ' }{ \varpi } h_1 -{\cal Y }  
\right).
\label{erq1}
\end{equation}
In vacuum and in the absence of a net electrical charge, the solution to
Maxwell equations is determined by the following set of differential
equations
\begin{equation}
{\cal H} = {\cal G} = {\cal Q} = 0 \ ,
\end{equation}
together with the regularity conditions ${\cal X} = {\cal Y} =0 $.
It is apparent that this is an overdetermined system if $ \varpi ' h_1 /
\varpi \ne 0$, for in this case the two equations $ {\cal Q} = 0 $ and $
{\cal Y} = 0 $ cannot be satisfied in the presence of
spacetime dragging.

An additional constraint $h_1 =0 $ leads to a trivial solution, $ g_n
=h_n =0$ and it is therefore impossible to impose $h_1 =0 $ everywhere.
As a result, the solutions found with this method are only approximate
solutions to the Maxwell equations in vacuum, with the deviation being
measured through the term $\varpi ' h_1 / \varpi $. If the value of this
quantity, which clearly depends on position, is small enough, then the
level of approximation can be very good. Hereafter we will monitor the
term $\varpi ' h_1 / \varpi $
as a consistency check to the solutions found.

The functions $g_1$ and $h_1$ are obtained from $ {\cal G} = {\cal X}
=0$.
\begin{eqnarray}
g_1  &=&  \frac{ \varpi r^2 }{2} h_0 '
%
%
 + \frac{ 3 \mu }{2 r^3 } 
\left[
 c_1 
   \left\{  3r^3 (r^2-3Mr+2M^2) \ln \alpha ^2 
+ 2M r^2 (3 r^2-6Mr+M^2) \right\}
 +\frac{J}{M}   
\right],
\label{solg1}
\\
  h_1 & =&
    - \frac{\varpi r^2}{2} g_0 ' 
\label{solg0}
\\
  & = &  \frac{ \mu \varpi  }{4 r}
\left[
 c_1   \left\{    6 r^3 ( 2r-3M ) \ln \alpha ^2  
+4Mr (6 r^2-3M r-M^2) \right\}
+ \frac{3J}{M^3} ( r \ln  \alpha ^2  +2M)
\right],
\label{solh1}
\end{eqnarray}
where a constant $c_1$ is given by the values at surface ($r=R$) such as
$ \alpha  _R = \alpha (R) $, 
\begin{equation}
c_1 = - \frac{ 1}{4 R M^3} \frac{ (R^2 \ln \alpha ^2 _R +2MR +2M^2 ) R^2
\Omega -2J( R \ln \alpha ^2 _R +2M ) } { 3R^2 (R-M) \alpha ^2 _R \ln
\alpha ^2 _R +6MR^2-12M^2R+2M^3 }.
\end{equation}
We have assumed that the stellar surface behaves as a perfect conductor
so that the following boundary condition can be imposed at the stellar
surface:
$  E_\theta + B_r \alpha ^{-1} \varpi R  \sin \theta  = 0 .$
Note that these analytic solutions have already been derived elsewhere
(e.g. \citet{k3,r1,r2}).
By direct calculations, we can check that these solutions satisfy $ {\cal
Q} =0, $ but never $ {\cal Y} = 0 $.  Indeed, these solutions are such
that ${\cal Y} = \varpi ' h_1 / \varpi \ne 0$ and for $r \gg M$ this term
can be easily estimated  to be 

\begin{equation}
\frac{  \varpi ' }{\varpi  } h_1 
 \sim  - \frac{ 9  }{10 }
\left(   4 M^6 c_1 + 5  J   \right)
\frac{ \mu \omega }{ M r^3 }
~~
\propto r^{-6} .
\label{erroh1}
\end{equation}
Since this term rapidly decreases with distance, the approximate
solutions will be rather accurate at a sufficient distance from the
stellar surface and could therefore be used in practical applications.

The analytic expressions (\ref{solh0}), (\ref{solg1}), (\ref{solg0}) and
(\ref{solh1}) are the same ones as obtained in previous works, although
with some slight differences in notations. Note that the ``error term''
(\ref{erroh1}) is effectively a {\it second-order} term in $\Omega$ and
is therefore absent in previous analyses which were restricted to the
low-frequency approximation. Because in our treatment such a term is not
neglected {\it a priori}, we can explicitly check the accuracy of the
expressions (\ref{solh0}), (\ref{solg1}), (\ref{solg0}) and (\ref{solh1})
at all radial positions.
Indeed, these solutions represent a rather good approximation and can be
used in practice in a slowly rotating spacetime, if the ``error term''
(which is usually quite small) is within a tolerable range.
%

\section{ Non-axially symmetric case}

   In the previous Section, the axis of the magnetic dipole is aligned
with the rotational axis. Here, instead, we consider what
happens when the dipole is misaligned. 
In particular, we focus on the case in which the axis of the magnetic 
dipole is perpendicular to the rotational axis.  

	The method used here to solve the Maxwell equations is almost
identical to the one used in the previous Section for the axially
symmetric configuration. In order to point out the
similarities, we will adopt the same notation for the radial functions $
g_n, h _n$. While this may induce some confusion, it is sufficient to
bear in mind that the basis functions used in the expansion are different
in the two cases [cf. eqs. (\ref{expndm0G})--(\ref{expndm0H}) and
(\ref{expndm1G})--(\ref{expndm1H})].

	Since the potentials $G$ and $H$ would depend on the azimuthal
wave number $m=1 $, they are expanded by the spherical harmonics
$Y_{2l}^1 $ and $Y_{2l+1} ^1 $, respectively. 
As a result, the angular expansion can be written as
\begin{equation}
\label{expndm1G}
 G(r, \theta, \phi, t) =e^{i\phi '} \sin \theta \cos \theta 
\left( \sum _{n=0} ^{\infty}
 g_n(r) (\sin \theta)^{2n}
\right),
\end{equation}
\begin{equation}
\label{expndm1H}
 H(r, \theta, \phi, t) =e^{i\phi '} \sin \theta \left(
\sum _{n=0} ^{\infty}
 h_n(r) (\sin \theta)^{2n} 
\right) ,
\end{equation}
where
\begin{equation}
\phi ' = \phi -\Omega t .
\end{equation}
The functions should depend only on $r, \theta, \phi ' $ from the
stationary condition as mentioned in Section 2.
As shown in eqs.(\ref{evct}) and (\ref{bvct}), the denominators in the
poloidal electric and magnetic fields become zero at the light cylinder.
Using the expansion forms,
we have two regularity conditions, 
which should be satisfied for $ r \ge r_c$.
The conditions for the radial functions
are written as 
\begin{equation}
\sum _{n=0} ^{\infty}
\left\{ 
 g_{n} ' + (2n +1)  \frac{ \varpi }{\alpha^2 } 
h_{n} \right\} b^{-2n}  = 0,
\end{equation}
\begin{equation}
\sum _{n=0} ^{\infty}
\left\{
 h_{n} '   
- \frac{ \varpi}{\alpha^2 } 
\left( 2n+1 - \frac{2n+2}{b^2}  \right)  g_{n} 
\right\} b^{-2n}
=0.
\end{equation}

%
  In a spherically symmetric and static spacetime the electromagnetic
fields are described completely in terms of four radial functions 
$ g_n, h_n (n=0,1).$  This set of functions can also provide 
approximate expressions in the case of a slowly rotating spacetime.
The regularity conditions for the non-vanishing components are given by
\begin{equation}
 0 ={\cal X}
\equiv 
 g_1 ' +b^2  g_0 '+\frac{\varpi b^2}{\alpha^2}h_0  +
\frac{3 \varpi}{\alpha^2} h_1 ,
%
%
\end{equation}
\begin{equation}
 0 = {\cal Y} 
\equiv
 h_1 ' + b^2  h_0 ' -  \frac{\varpi}{\alpha^2}(b^2 -2) g_0  
- \frac{\varpi}{\alpha^2} \left( 3-\frac{4}{b^2}\right)  g_1 .
%
\end{equation}
Using these conditions,
the expressions for the electromagnetic fields
regular at the light cylinder are
\begin{equation}
B_r = 
 - \left[
 h_0 ' - \frac{\varpi}{\alpha^2}  g_0  -
 \frac{4}{ \varpi r^2} g_1 \sin ^2 \theta  
  \right]
e^{i\phi '} \sin \theta ,
\end{equation}
\begin{equation}
B_\theta
 = - \frac{1}{\alpha r}
 \left[  h_0 
 +( b^2 h_0 +3 h_1 +\varpi r^2 g_0 ')\sin ^2 \theta 
 \right] e^{i\phi '} \cos \theta ,
\end{equation}
\begin{equation}
B_\phi
 = -\frac{i}{\alpha r}
\left[  h_0+ h_1 \sin ^2 \theta  \right] e^{i\phi '}  ,
\end{equation}
\begin{equation}
E_r =  - \left[
 g_0 ' + \frac{\varpi}{\alpha^2} h_0  
\right] 
e^{i\phi '} \sin \theta  \cos \theta ,
\end{equation}
\begin{equation}
E_\theta  = - \frac{1}{\alpha r}
  \left[ g_0  - \left\{
 (2-b^2) g_0-3 g_1 +\varpi  r^2 h_0 ' \right\} \sin ^2 \theta 
  \right] e^{i\phi '} ,
\end{equation}
\begin{equation}
E_\phi  = -\frac{i}{\alpha r}
  \left[ g_0 +  g_1 \sin ^2 \theta  \right] e^{i\phi '} \cos \theta .
\end{equation}
The divergence of the electric and magnetic fields yields
\begin{equation}
\Div \EV  = - \alpha {\cal G }  \sin \theta \cos \theta ,
\end{equation}
\begin{equation}
\Div \BV  = - \alpha {\cal H } \sin \theta
- {\cal Q }  \sin^3 \theta , 
\end{equation}
where 
\begin{eqnarray}
{\cal G } & =&
g_0 ''+\frac{2}{r} g_0 '-
\frac{2 \varpi}{ \alpha ^2} h_0 '
+ \frac{8}{ (\alpha r)^2 } g_1
%
%
+\frac{1}{(\alpha r)^2}(3 b^2 -6 ) g_0
+ \frac{1 }{r^2} 
\left( \frac{ \varpi r^2 }{\alpha^2} \right) ' 
 h_0 ,
\label{obl0G00}
\\
{\cal H }  &=&
h_0 ''+\frac{2}{r} h_0 '
+ \frac{2 \varpi}{ \alpha ^2} g_0 ' 
- \frac{1 }{r^2} 
\left( \frac{ \varpi r^2 }{\alpha^2} \right) ' 
g_0
%
%
+\frac{8}{ (\alpha r)^2 } h_1+
\frac{1}{(\alpha r)^2}(3 b^2 -2) h_0 ,
\label{obl0H00}
\\
{\cal Q } &=&
\frac{ 4 \alpha }{ \varpi r^2}
\left( \frac{\varpi ' }{ \varpi } g_1 - {\cal X }
\right) .
\label{obq1}
\end{eqnarray}
The solutions $g_n, h_n (n=0,1)$ of the Maxwell equations in vacuum are
determined by $ {\cal G }= {\cal H }= {\cal Q } =0 $ and the regularity
conditions $ {\cal X }= {\cal Y }= 0 $, but it is clear that the system
of differential equations is overdetermined in a slowly rotating
spacetime because of the presence of a redundant equation. In this
respect, and as mentioned in the previous Section, the solutions to these
equations are not exact.

	In order to obtain the Deutsch-type solution, we impose the
following conditions (see the Appendix for details)
\begin{equation}
  g_1 = 0, 
~~
  h_1 = - \frac{1}{3} 
\left( \varpi r^2 g_0 '  + b^2  h_0
\right).
\label{cndgh1} 
\end{equation}
Using these conditions, we have ${\cal X } = 0$ and ${\cal Q } =0 $.
The two functions $ g_0, h_0 $ are determined by solving $ {\cal G }
={\cal H } = 0$. We introduce the
functions $F_{l} (l=1,2 )$ so as to express $h_0$ and $g_0$ as
\begin{equation}
  g_0 = \frac{\alpha^2}{6} F_{2} '
- \frac{\varpi}{2} F_{1} ,
\label{ndfg0}
\end{equation}
\begin{equation}
  h_0 = \frac{\alpha^2}{2}  F_{1} '
+ \frac{\varpi}{6} F_{2},
\label{ndfh0}
\end{equation}
where $F_{1} $ and $F_{2} $ satisfy the coupled equations
\begin{equation}
\alpha^2 ( \alpha^2 F_{1} ^{\prime} )^{\prime}
 + \left( \varpi^2  -
      \frac{2 \alpha^2}{r^2}  \right)  F_{1} 
+ \frac{1}{3} \alpha^2 \varpi ' F_{2} =0,
\label{eqf1}
\end{equation}
\begin{equation}
\alpha^2 ( \alpha^2 F_{2} ^{\prime} )^{\prime}
 + \left( \varpi^2  -
      \frac{6 \alpha^2}{r^2}  \right)  F_{2} 
- 3 \alpha^2 \varpi '  F_{1} =0.
\label{eqf2}
\end{equation}
It is easy to check that the functions (\ref{ndfg0}) and (\ref{ndfh0})
satisfy the coupled equations ${\cal G } ={\cal H } =0 $ with the
conditions (\ref{cndgh1}).
Indeed, in the limit of a non-rotating spacetime, equations (\ref{eqf1})
and (\ref{eqf2}) become the Regge-Wheeler wave-like equations with
frequency $\Omega$ for $ l=1,2 $ \citep{w1}.
If the spacetime is rotating, however, the two equations are coupled
because of the spacetime dragging through the coefficient $ \varpi' = -
\omega' =6J/r^4$ .
From this choice of functions, we have
\begin{equation}
{\cal Y} =   - \frac{ 2J }{r^4}  F_{2} .
\label{ery1}
\end{equation}
This does not vanish in general, but rapidly decreases to zero with the
radius because of the factor $ 2J/ r^4 $.
In fact, we can find 
${\cal Y}  \sim - 6 \mu J \Omega R^2 /r^6 $
near the light cylinder and
${\cal Y}  \sim  2 \mu J \Omega ^3  R^2 
\exp \{ i\Omega(r +2M \ln(r/2M -1))\} /r^4 $  
beyond the light cylinder.

	Since the function $F_2 $ describes the induced electromagnetic
fields by a rotating perfect conductor, it is a first-order quantity in
$\Omega$. Consequently, the right-hand-side of (\ref{ery1}) is
second-order in $\Omega$ and was thus absent in previous works performed
in the slow-frequency approximation.  Furthermore, eqs. (\ref{eqf1}) and
(\ref{eqf2}) are significantly simplified when the terms ${\cal
O}(\Omega^2)$ can be neglected and, in fact, analytic expressions have
been found in this case by \citet{r1,r2}. It should be stressed, however,
that such analytic solutions can be valid only in the vacuum exterior
close to the stellar surface, i.e. for $r <r_c$.  Beyond the light
cylinder, in fact, the wave-like nature of the electromagnetic fields
becomes important and the term ${\cal O}(\Omega^2)$ in eqs.  (\ref{eqf1})
and (\ref{eqf2}) can no longer be neglected.  

Expressing the electromagnetic fields in terms of functions $F_l$
we have
\begin{eqnarray}
\vec{B}&=&
\left[
-\frac{1}{r^2} F_1 \sin \theta e^{i \phi'} ,~~
-\left( 
\frac{\alpha}{2r} F_1  ' + \frac{\varpi}{6 \alpha r} F_2
 \right) \cos \theta e^{i \phi'} ,~~
- i \left( 
\frac{\alpha}{2r} F_1  ' + \frac{\varpi}{6 \alpha r} F_2
 \cos 2 \theta \right) e^{i \phi'} 
\right] ,
\\
\vec{E}&=&
\left[
- \frac{1}{2r^2} F_2 \sin 2 \theta e^{i \phi'} , ~~
\left( 
   \frac{\varpi}{2 \alpha r} F_1
 - \frac{\alpha}{6r} F_2  ' \cos 2 \theta 
 \right) e^{i \phi'} ,~~
 i \left( 
  \frac{\varpi}{2 \alpha r} F_1
 - \frac{\alpha}{6r} F_2  ' 
\right) \cos \theta  e^{i \phi'} 
\right] ,
\end{eqnarray}
and we can use the perfect conductor condition at the surface, i.e.  $
E_\theta + B_r \alpha ^{-1} \varpi R \sin \theta = 0 ,$ which is reduced
to $ F_2 ' -3 \varpi F_1 /\alpha ^{2}= 0 $ at $ r=R.$

	Using this boundary condition, the numerical solution of the
coupled equations (\ref{eqf1}) and (\ref{eqf2}) for the magnetic field
$B_r$ is shown in Fig.1 for a typical choice of the stellar parameters.
In the same figure, we also show the approximate solution in curved
spacetime by \citet{r1,r2} [indicated by (a)]
and the asymptotic one in flat spacetime by
\citet{d1} [indicated by (f)].
The function $B_r$ decreases with radius and the difference between our
numerical result and analytic expressions is not so clear when shown in
absolute magnitude. To highlight the differences we have therefore
multiplied the magnetic field intensity $B_r$ in Fig.1 by the factor $
r^3 /(\mu \sin \theta e^{i \phi'})$.
Note that our numerical solution agrees rather well with that of
\citet{r1,r2} near the surface. Around $ r \sim 0.2 \Omega ^{-1}$, our
result deviates from it, approaching however the solution for flat
spacetime by \citet{d1} as the distance from the stellar surface
increases. 

	The explanation for this behavior is rather simple since the
analytic expression for curved spacetime in the low-frequency
approximation is valid only well within the light cylinder region $ r
\sim 0.2 \Omega^{-1} < r_c$, where the relativistic corrections are large
and the electromagnetic frequencies small. Indeed, in the low-frequency
approximation the point $ r_c \sim 1/ \Omega$ is regarded as ``infinity''
and, as expected, the results of the low-frequency approximation cannot
be accurate near this limit.
As the distance from the star increases, a qualitative difference appears
from the solutions of \citet{r1,r2} and this is represented by the
sinusoidal, wave-like nature which is clear beyond the light cylinder.
In this region, the gravitational corrections are not important and, in
fact, our numerical result agrees well with that of flat spacetime.
Fig.1  therefore serves to illustrate how our approach is very useful in
connecting two different regimes, i.e.  the relativistic one and the and
the asymptotically flat one within a single, semi-analytic approach.

\begin{figure}
%
\begin{minipage}{0.45\linewidth}
\centering
\includegraphics[scale=1.35]{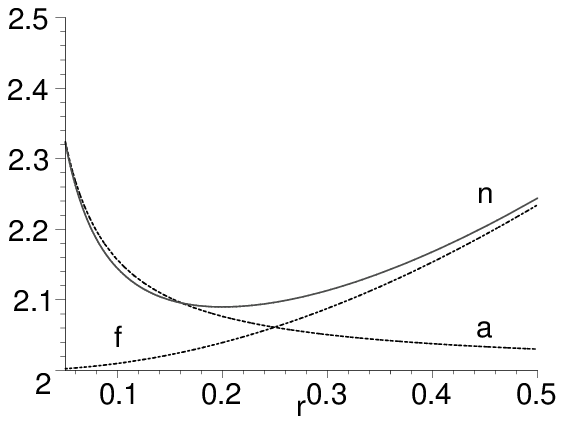}
\end{minipage}
\hspace{10mm}
\begin{minipage}{0.45\linewidth}
\centering
\includegraphics[scale=1.35]{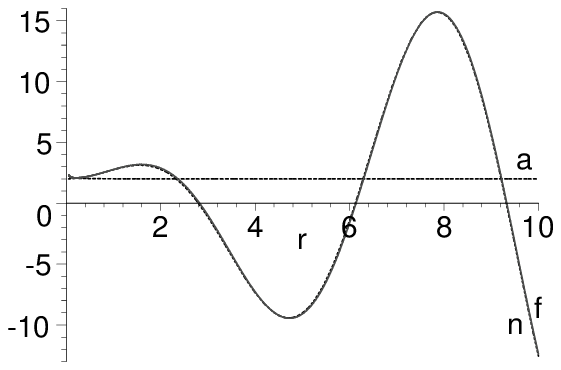}
\end{minipage}

\caption{
Magnetic field $B_r r^3 /(\mu \sin \theta  e^{i \phi'})$
as a function of radius $r/r_c$.
The left and right panels show the behavior near the surface
$(  R \le r \le r_c/2 )$  
and that for wave region
$(  R \le r \le 10 r_c ).$   
The solid line with symbol 'n' represents our result
solved for $2M/r_c=0.01, R/r_c=0.05, J=10^{-3} r_c ^2$.
The dotted lines with symbols 'a' and 'f'
respectively represent the analytic result for curved spacetime
\citep{r1,r2} and that for flat spacetime \citep{d1}.
}
\end{figure}

\section{ Concluding Remarks}
	We have studied the stationary solutions of the Maxwell equations
in slowly rotating spacetime. The electromagnetic fields exterior to
rotating magnetic dipole moment have been derived with the perfect
conductor condition at the surface. This problem is not new, but it has
been here considered through a different type of approach and with a
different level of accuracy. In particular, we have considered the
general relativistic electromagnetic fields exterior to a rotating
neutron star in the slow-rotation approximation but {\it not} in the
low-frequency approximation. The solutions derived here reduce to the
exact, analytic ones if the spacetime is spherically symmetric and no
frame dragging effect is present. Furthermore, in this same limit, they
can be expressed in terms of analytic functions
in the limit of flat Minkowski space \citep{d1}.

	Using these results as a guide, we have looked for solutions that
have similar angular dependence but more complicated radial functional
behavior. Adopting a semi-analytical approach in which simple
expressions for the electromagnetic fields need to be completed through
the calculation of coefficients coming from the numerical solution of two
partial differential equations, we have found approximate solutions to
the Maxwell equations. The error terms present in our solutions are
${\cal O}(\Omega^2)$ and were therefore absent in previous approaches
using the low-frequency approximation. The numerical solutions found in
this way agree well with the known analytic results both in the
strong-field region \citep{r1,r2} and in the asymptotically flat one
\citep{d1}, where the wave nature of the solution is important beyond the
light cylinder in the case of non-axially symmetric
configurations. Furthermore, our numerical solutions represent a useful
tool to determine when these two analytic approaches cease to be
accurate. Overall, the new method proposed here offers the advantage of
being able to bridge in a single approximate but accurate formalism both
the strongly relativistic and the asymptotically flat regimes.

	Our approach is fundamentally based on a series expansion for
which we discuss here only the lowest-order terms. However, it is in
principle possible to increase the accuracy of our results by increasing
the number of expansion series.  For example, by including
the higher-order functions 
$g_2, h_2$, the solutions can be calculated for a larger system of
partial differential equations. Also in this case, there would be a
redundant equation in the set of equations signalling terms which are
incompatible within a set of differential equations. While we will not
show any explicit calculations here, it is not difficult to realize that
in the lowest extension to the results presented here the new correction
would come through a term $\sim \varpi ' h_2 /\varpi $
(or $ \varpi ' g_2 /\varpi $)  in the system of
the functions $ g_n, h_n(n=0,1,2)$. The function $h_2$ (or $g_2$)
is expected to be
small, because it should be zero both at the surface and infinity and the
overall accuracy is therefore improved in this enlarged system of
equations.
By increasing further the number of functions, the difference between the
incompatible equations becomes small. Of course, the limit of $ n \gg 1$
may be preferable because of its increase in accuracy, but it also has
clear limitations in practical applications.
%

\section*{Acknowledgments}
We thank the referee, Luciano Rezzolla,
for many useful comments that improved the manuscript.
This work was supported in part
by a Grant-in-Aid for Scientific Research 
(No.~14047215) from 
the Japanese Ministry of Education, Culture, Sports,
Science and Technology.
%



\appendix
\section[]{Deutsch Solution}

	We here summarize the Deutsch solution and offer it for
comparison. The solution describes the electromagnetic fields produced by
a dipole magnet rotating with angular frequency $\Omega$.
	The quadrupolar electric field is induced by a perfect-conductor
condition at the surface, written as $ E_\theta + \Omega R \sin \theta
B_r=0$ at $r=R.$
We here assume that the angle between the dipole moment and rotational
axis is $\chi$ so that the components of the electromagnetic fields are
given by
\begin{eqnarray}
 B_r  &=& 
 \frac{2  \mu}{r^3} \cos \chi \cos \theta 
+ \frac{2  \mu}{r^3} s_1 e^{i \lambda}
 \sin \chi \sin \theta ,
\\
 B_\theta  &=& 
 \frac{\mu }{r^3}  \cos \chi \sin \theta 
- \frac{ \mu}{r^3}  ( s_2  -d_1 s_3 ) e^{i \lambda}
 \sin \chi \cos \theta ,
\\
 B_\phi  &=&
-\frac{ i \mu }{r^3}
 \left(  s_2  - d_1 s_3   \cos 2 \theta \right) e^{i \lambda}\sin \chi,
\\
 E_r 
&=& 
- \frac{2\mu \Omega R^2}{r^4}   \cos \chi P_2 (\theta)
+\frac{ 3 \mu  d_1 }{ \Omega  r^4} s_3  e^{i \lambda}
 \sin \chi \sin 2 \theta ,
\\
%
 E_\theta 
&=& 
-  \frac{2 \mu\Omega R^2}{r^4}   \cos \chi \sin 2 \theta 
 -\left(
 \frac{  \mu  \Omega }{r^2} s_1
+\frac{ 2 \mu  d_1 }{ \Omega  r^4} s_4 \cos 2 \theta
\right) \sin \chi    e^{i \lambda}  ,
\\
%
 E_\phi 
&=&   -i \left(
 \frac{ \mu  \Omega }{r^2} s_1
+\frac{2 \mu d_1 }{ \Omega r^4} s_4
 \right) \sin \chi  \cos \theta  e^{i \lambda}, 
\end{eqnarray}
where
\begin{eqnarray}
  s_1 &=& 1 -i \Omega r ,
\\
  s_2 &=& 1 -i \Omega r  - \Omega^2 r^2 , 
\\
  s_3 &=& 
 1 -i \Omega r  - \frac{1}{3} \Omega^2 r^2 , 
\\
  s_4 &=& 
 1 -i \Omega r  - \frac{1}{2} \Omega^2 r^2 
+ \frac{i}{6} \Omega^3 r^3   ,
\\
  \lambda &=&  \phi -\Omega (t- r) ,
\\
d_1 & =& - \frac{ 3 \Omega ^2 R^2 (1-i \Omega R  )} 
{ 6 -i 6 \Omega R -3 \Omega ^2 R^2 +i \Omega ^3 R^3 } .
\end{eqnarray}

\label{lastpage}

\begin{thebibliography}{99}
%
\bibitem[\protect\citeauthoryear{Anderson \&  Cohen }{1970}]{a1} 
 Anderson J.L.,  Cohen J.M.,  1970,
Ap \&SS, 9, 146
%
\bibitem[\protect\citeauthoryear{Bardeen et al.}{1972}]{b1} 
 Bardeen J.M., Press W.H., Teukolsky S.A., 1972, 
ApJ, 178, 342
%
\bibitem[\protect\citeauthoryear{Deutsch}{1955}]{d1} 
 Deutsch A.J., 1955,
Annales d'astrophysique, 18, 1
%
\bibitem[\protect\citeauthoryear{Ginzburg \&  Ozernoy }{1964}]{g1} 
 Ginzburg V.L.,  Ozernoy  L.M., 1964,
Zh. Zksp.Teor. Fiz,  47, 1030
%
\bibitem[\protect\citeauthoryear{Konno et al.}{1999}]{k1} 
 Konno K.,  Obata T.  Kojima Y., 1999,
A \& AS,  352, 211
%
\bibitem[\protect\citeauthoryear{Konno et al.}{2000}]{k2} 
 Konno K.,  Obata T.  Kojima Y.,  2000,
A \& AS, 356, 234
%
\bibitem[\protect\citeauthoryear{Konno \& Kojima}{2000}]{k3} 
 Konno K.,  Kojima Y., 2000, 
Prog. Theor. Phys., 104, 1117
%
\bibitem[\protect\citeauthoryear{Kouveliotou et al.}{1998}]{ok1} 
 Kouveliotou C.,  Dieters S.,  Strohmayer T.,  et al. 1998,
Nat, 393, 235
%
\bibitem[\protect\citeauthoryear{Kouveliotou et al.}{1999}]{ok2} 
 Kouveliotou C.,  Strohmayer T., Hurley K.,  et al. 1999,
ApJ, 510, L115
%
\bibitem[\protect\citeauthoryear{Mereghetti  \& Stella}{1995}]{ms1} 
 Mereghetti S.,  Stella L., 1995,
ApJ, 442, L17
%
\bibitem[\protect\citeauthoryear{Muslimov \& Tysgan }{1992}]{m1} 
 Muslimov A.,  Tysgan A.I.,  1992,
MNRAS,  255,  61
%
\bibitem[\protect\citeauthoryear{Muslimov \& Harding }{1997}]{m2} 
 Muslimov A.,  Harding A.K., 1997,
ApJ, 485, 735
%
\bibitem[\protect\citeauthoryear{Parker}{1979}]{bp1} 
 Parker E.N., 1979, 
{\it Cosmical Magnetic Fields.} 
Claredon Press, Oxford 
%
\bibitem[\protect\citeauthoryear{Petterson }{1974}]{p1} 
 Petterson J.A., 1974,
Phys.Rev. D10, 3166
%
%
\bibitem[\protect\citeauthoryear{Thompson \& Duncan}{1993}]{td1} 
 Thompson C., Duncan R.C., 1993,
ApJ 408, 194
%
\bibitem[\protect\citeauthoryear{Thompson \& Duncan}{1995}]{td2} 
 Thompson C., Duncan R.C., 1995,
MNRAS,  275,  255
%
\bibitem[\protect\citeauthoryear{Thompson \& Duncan}{1996}]{td3} 
 Thompson C., Duncan R.C., 1996,
ApJ, 473, 322
%
\bibitem[\protect\citeauthoryear{Thorne et al.}{1986}]{t1} 
 Thorne K.S., Price R.H., Macdonald D.A., 1986, 
{\it Black Hole: The Membrane Paradigm.} 
Yale University Press, New Haven
%
\bibitem[\protect\citeauthoryear{Rezzolla et al.}{2001a}]{r1} 
 Rezzolla L., Ahmedov B.J.,  Miller J.C., 2001a, 
MNRAS, 322, 723
%
\bibitem[\protect\citeauthoryear{Rezzolla et al.}{2001b}]{r2} 
 Rezzolla L., Ahmedov B.J.,  Miller J.C., 2001b,
Found. Phys., 31, 1051
%
\bibitem[\protect\citeauthoryear{Wheeler}{1955}]{w1} 
 Wheeler. J.A., 1955, 
Phys. Rev., 97, 511
%
\bibitem[\protect\citeauthoryear{Zanotti \&  Rezzolla}{2002}]{z1} 
 Zanotti O,  Rezzolla L., 2002, 
MNRAS, 331, 376
%
\bibitem[\protect\citeauthoryear{Zeldovich et al.}{1983}]{bz1} 
 Zeldovich Ya.B.,  Ruzmaikin A.A.,  Sokoloff D.D., 1983, 
{\it Magnetic Fields in Astrophysics.} 
Gordon and Breach Science Publishers, New York
\end{thebibliography}
               \end{document}